\documentclass[twocolumn]{article}
\usepackage{graphicx}
\usepackage{dcj}
\usepackage{authblk}
\usepackage{url}

\begin{document}

\title{Compression of next-generation sequencing reads aided by highly efficient de novo assembly}

\author[1]{Daniel C. Jones\thanks{dcjones@cs.washington.edu}}
\author[1,2,3]{Walter L. Ruzzo\thanks{ruzzo@cs.washington.edu}}
\author[4]{Xinxia Peng\thanks{xinxiap@u.washington.edu}}
\author[4]{Michael G. Katze\thanks{honey@u.washington.edu}}
\affil[1]{Department of Computer Science and Engineering, University of Washington}
\affil[2]{Department of Genome Sciences, University of Washington}
\affil[3]{Fred Hutchinson Cancer Research Center}
\affil[4]{Department of Microbiology, University of Washington}

\maketitle

\begin{abstract}
We present Quip, a lossless compression algorithm for next-generation
sequencing data in the FASTQ and SAM/BAM formats. In addition to implementing
reference-based compression, we have developed, to our knowledge, the first
assembly-based compressor, using a novel de novo assembly algorithm. A
probabilistic data structure is used to dramatically reduce the memory
required by traditional de Bruijn graph assemblers, allowing millions of reads
to be assembled very efficiently. Read sequences are then stored as positions
within the assembled contigs. This is combined with statistical compression of
read identifiers, quality scores, alignment information, and sequences,
effectively collapsing very large datasets to less than 15\% of their original
size with no loss of information.

Availability: Quip is freely available under the BSD license from
\url{http://cs.washington.edu/homes/dcjones/quip}.
\end{abstract}

\section{Introduction}

With the development of next-generation sequencing (NGS) technology
researchers have had to adapt quickly to cope with the vast increase in raw
data. Experiments that would previously have been conducted with microarrays
and resulted in several megabytes of data, are now performed by sequencing,
producing many gigabytes, and demanding a significant investment in
computational infrastructure. While the cost of disk storage has also steadily
decreased over time, it has not matched the dramatic change in the cost and
volume of sequencing. A transformative breakthrough in storage technology may
occur in the coming years, but the era of the \$1000 genome is certain to
arrive before that of the \$100 petabyte hard disk.

As cloud computing and software as a service become increasingly relevant to
molecular biology research, hours spent transferring NGS datasets to and from
off-site servers for analysis will delay meaningful results. More often
researchers will be forced to maximize bandwidth by physically transporting
storage media (via the ``sneakernet''), an expensive and logistically
complicated option. These difficulties will only be amplified as exponentially
more sequencing data is generated,  implying that even moderate gains in
domain-specific compression methods will translate into a significant
reduction in the cost of managing these massive datasets over time.

\enlargethispage{-65.1pt}

Storage and analysis of NGS data centers primarily around two formats that
have arisen recently as de facto standards: FASTQ and SAM. FASTQ stores, in
addition to nucleotide sequences, a unique identifier for each read and
quality scores, which encode estimates of the probability that each base is
correctly called. For its simplicity, FASTQ is a surprisingly ill-defined
format. The closest thing to an accepted specification is the description by
\citet{Cock2010}, but the format arose ad hoc from multiple sources (primarily
Sanger and Solexa/Illumina), so a number of variations exist, particularly in
how quality scores are encoded. The SAM format is far more complex but also
more tightly defined, and comes with a reference implementation in the form of
SAMtools \citep{Li2009b}. It is able to store alignment information in
addition to read identifiers, sequences, and quality scores. SAM files, which
are stored in plain text, can also be converted to the BAM format, a
compressed binary version of SAM, which is far more compact and allows for
relatively efficient random access.

Compression of nucleotide sequences has been the target of some interest, but
compressing NGS data, made up of millions of short fragments of a greater
whole, combined with metadata in the form of read identifiers and quality
scores, presents a very different problem and demands new techniques.
Splitting the data into separate contexts for read identifiers, sequences, and
quality scores and compressing them with the Lempel-Zip algorithm and Huffman
coding has been explored explored by \citet{Tembe2010} and
\citet{Deorowicz2011}, who demonstrate the promise of domain-specific
compression with significant gains over general-purpose programs like gzip and
bzip2.

\citet{Kozanitis2011} and \citet{Hsi-YangFritz2011} proposed reference-based
compression methods, exploiting the redundant nature of the data by aligning
reads to a known reference genome sequence and storing genomic positions in
place of nucleotide sequences. Decompression is then performed by copying the
read sequences from the genome. Though any differences from the reference
sequence must also be stored, referenced-based approaches can achieve much
higher compression and they grow increasing efficient with longer read
lengths, since storing a genomic position requires the same amount of space,
regardless of the length of the read.

This idea is explored also in the Goby format
(\url{http://campagnelab.org/software/goby/}), which has been
proposed an alternative to SAM/BAM, the primary functional difference being
that sequences of aligned reads are not stored but looked up in a reference
genome when needed (frequently they are not). For some applications,
reference-based compression can be taken much further by storing only SNP
information, summarizing a sequencing experiment in several megabytes
\citep{Christley2009}. However, even when SNP calls are all that is needed,
discarding the raw reads would prevent any reanalysis of the data.

While a reference-based approach typically results in superior compression it
has a number of disadvantages. Most evidently, an appropriate reference
sequence database is not always available, particularly in the case of
metagenomic sequencing. One could be contrived by compiling a set of genomes
from species expected to be represented in the sample. However, a high degree
of expertise is required to curate and manage such a project-dependent
database. Secondly, there is the practical concern that files compressed with
a reference-based approach are not self-contained. Decompression requires
precisely the same reference database used for compression, and if it is lost
or forgotten the compressed data becomes inaccessible.


Another recurring theme in the the growing literature on short read
compression is lossy encoding of sequence quality scores. This follows
naturally from the realization that quality scores are particularly difficult
to compress. Unlike read identifiers, which are highly redundant, or
nucleotide sequences, which contain some structure, quality scores are
inconsistently encoded between protocols and computational pipelines and are
often simply high-entropy. It is dissatisfying that metadata (quality scores)
should consume more space than primary data (nucleotide sequences). Yet, also
dissatisfying to many researchers is the thought of discarding information
without a very good understanding of its effect on downstream analysis.

A number of lossy compression algorithms for quality scores have been
proposed, including various binning schemes implemented in QScores-Archiver
\citep{Wan2011} and SCALCE (\url{http://scalce.sourceforge.net}), scaling to a
reduced alphabet with randomized rounding in SlimGene \citep{Kozanitis2011}, and
discarding quality scores for bases which match a reference sequence
in Cramtools
\citep{Hsi-YangFritz2011}. In SCALCE, SlimGene, and Cramtools, quality scores
may also be losslessly compressed. \citet{Kozanitis2011} analyzed of the effects of their
algorithm on downstream analysis. Their results suggest that while some SNP
calls are affected, they are primarily marginal, low-confidence calls between
hetero- and homozygosity.

Decreasing the entropy of quality scores while retaining accuracy is an
important goal, but successful lossy compression demands an understanding of
what is lost. For example, lossy audio compression (e.g. MP3) is grounded in
psychoacoustic principles, preferentially discarding the least perceptible
sound. Conjuring a similarly principled method for NGS quality scores is
difficult given that both the algorithms that generate them and the algorithms
that are informed by them are moving targets. In the analysis by
\citet{Kozanitis2011}, the authors are appropriately cautious in interpreting
their results, pointing out that ``there are dozens of downstream applications
and much work needs to be done to ensure that coarsely quantized quality
values will be acceptable for users.''


In the following sections we describe and evaluate Quip, a new lossless
compression algorithm which leverages a variety of techniques to achieve very
high compression over sequencing data of many types, yet remains efficient and
practical. We have implement this approach in a open-source tool that is
capable of compressing both BAM/SAM and FASTQ files, retaining all information
from the original file.

\section{Materials \& Methods}

\subsection{Statistical Compression}

The basis of our approach is founded on statistical compression using
arithmetic coding, a form of entropy coding which approaches optimality, but
requires some care to implement efficiently (see \citet{Said2004} for an
excellent review). Arithmetic coding can be thought of as a refinement of
Huffman coding, the major advantage being that it is able to assign codes of a
non-integral number of bits. If a symbol appears with probability 0.1, it can
be encoded near to its optimal code length of $-\log_2(0.1) \approx 3.3$ bits.
Despite its power, it has historically seen much less use than Huffman coding,
due in large part to fear of infringing on a number of patents that have now
expired.

Arithmetic coding is a particularly elegant means of compression in that it
allows a complete separation between statistical modeling and encoding. In
Quip, the same arithmetic coder is used to encode quality scores, read
identifiers, nucleotide sequences, and alignment information, but with very
different statistical models for each, which gives it a tremendous advantage
over general-purpose compression algorithms that lump everything into a
single context. Futhermore, all parts of our algorithm use adaptive modeling:
parameters are trained and updated as data is compressed, so that an
increasingly tight fit and higher compression is achieved on large files.

\subsubsection{Read Identifiers}

The only requirement of read identifiers is that they uniquely identify the
read. A single integer would do, but typically each read comes with a complex
string containing the instrument name, run identifier, flow cell identifier,
and tile coordinates. Much of this information is the same for every read and
is simply repeated, inflating the file size.

To remove this redundancy, we use a form of delta encoding. A  parser
tokenizes the ID into separate fields which are then compared to the previous
ID. Tokens that remain the same from read to read (e.g. instrument name)
can be compressed to a negligible amount of space --- arithmetic coding produces
codes of less than 1 bit in such cases. Numerical tokens are recognized and
stored efficiently, either directly or as an offset from the token in the same
position in previous read. Otherwise non-identical tokens are encoded by
matching as much of the prefix as possible to the previous read's token before
directly encoding the non-matching suffix.

The end result is that read IDs, which are often 50 bytes or longer, are
typically stored in 2-4 bytes. Notably, in reads produced from Illumina
instruments, most parts of the ID can be compressed to consume almost no
space; the remaining few bytes are accounted for by tile coordinates. These
coordinates are almost never needed in downstream analysis, so removing them
as a preprocessing step would shrink file sizes even further. The parser used
is suitably general so that no change to the compression algorithm would be
needed.

\subsubsection{Nucleotide Sequences}

To compress nucleotide sequences, we adopt a very simple model based on high-order
Markov chains. The nucleotide at a given position in a read is predicted
using the preceding twelve positions. This model uses more memory than
traditional general-purpose compression algorithms ($4^{13} = 67,108,864$
parameters are needed, each represented in 32 bits)  but it is simple and
extremely efficient (very little computation is required and run time is
limited primarily by memory latency, as lookups in such a large table result
in frequent cache misses).

An order-12 Markov chain also requires a very large amount of
data to train, but there is no shortage with the datasets we wish to
compress. Though less adept at compressing extremely short files,
after compressing several million reads the parameters are tightly
fit to the nucleotide composition of the dataset so that the remaining reads
will be highly compressed. Compressing larger files only results in a tighter
fit and higher compression.

\subsubsection{Quality Scores}

It has been previously noted that the quality score at a given position is
highly correlated with the score at the preceding position
\citep{Kozanitis2011}. This makes a Markov chain a natural model, but unlike
nucleotides, quality scores are over a much larger alphabet (typically 41--46
distinct scores). This limits the order of the Markov chain: long chains will
require a great deal of space and take a unrealistic amount of data to train.

To reduce the number of parameters, we use an order-3 Markov chain, but
coarsely bin the distal two positions. In addition to the preceding three
positions, we condition on the position within the read and a running count of
the number large jumps in quality scores between adjacent positions (where
a ``large jump'' is defined as $|q_{i} - q_{i-1}| > 1$), which allows reads
with highly variable quality scores to be encoded using separate models. Both
of these variables are binned to control the number of parameters.

\subsection{Reference-Based Compression}

We have also implemented lossless reference-based compression. Given aligned
reads in SAM or BAM format, and the reference sequence to which they are
aligned (in FASTA format), the reads are compressed preserving all information
in the SAM/BAM file, including the header, read IDs, alignment information,
and all optional fields allowed by the SAM format. Unaligned reads are
retained and compressed using the Markov chain model.

\subsection{Assembly-Based Compression}

To complement the reference-based approach, we developed an assembly-based
approach which offers some of the advantages of reference-based compression,
but requires no external sequence database and produces files which are
entirely self-contained. We use the first (by default) 2.5 million reads to
assemble contigs which are then used in place of a reference sequence
database to encode aligned reads compactly as positions.

Once contigs are assembled, read sequences are aligned using a simple ``seed
and extend'' method: 12-mer seeds are matched using a hash table, and
candidate alignments are evaluated using Hamming distance. The best
(lowest Hamming distance) alignment is chosen, assuming it falls below a given
cutoff, and the read is encoded as a position within the contig set. Roughly,
this can be thought of as a variation on the Lempel-Ziv algorithm: as
sequences are read, they are matched to previously observed data, or in this
case, contigs assembled from previously observed data. These contigs are
not explicitly stored, but rather reassembled during decompression.


Traditionally, de novo assembly is extremely computationally intensive. The
most commonly used technique involves constructing a de Bruijn graph, a
directed graph in which each vertex represents a nucleotide $k$-mer present in
the data for some fixed $k$ (e.g., $k = 25$ is a common choice). A directed
edge from a $k$-mer $u$ to $v$ occurs if and only if the $(k - 1)$-mer suffix
of $u$ is also the prefix of $v$. In principle, given such a graph, an
assembly can be produced by finding an Eulerian path, i.e., a path that
follows each edge in the graph exactly once \citep{Pevzner2001}. In practice,
since NGS data has a non-negligible error rate, assemblers augment each vertex
with the number of observed occurrences of the $k$-mer and leverage these
counts using a variety of heuristics to filter out spurious paths.

A significant bottleneck of the de Bruijn graph approach is building an
implicit representation of the graph by counting and storing $k$-mer
occurrences in a hash table. The assembler implemented in Quip 
overcomes this bottleneck to a large extent by using a data structure based on
the Bloom filter to count $k$-mers. The Bloom filter \citep{Bloom1970} is a
probabilistic data structure that represents a set of elements extremely
compactly, at the cost of elements occasionally colliding and incorrectly
being reported as present in the set. It is probabilistic in the sense that
these collisions occur pseudo-randomly, determined by the size of the table
and the hash functions chosen, but generally with low probability.

The Bloom filter is generalized in the counting Bloom filter, in which an
arbitrary count can be associated with each element \citep{Fan2000}, and
further refined in the d-left counting Bloom filter (dlCBF)
\citep{Bonomi2006}, which requires significantly less space than the already
quite space efficient counting Bloom filter. We base our assembler on a
realization of the dlCBF. Because we use a probabilistic data structure,
$k$-mers are occasionally reported to have incorrect (inflated) counts. The
assembly can be made less accurate by these incorrect counts, however a poor
assembly only results in slightly increasing the compression ratio.
Compression remains lossless regardless of the assembly quality, and in
practice collisions in the dlCBF occur at a very low rate (this is explored in
the results section).

Given a probabilistic de Bruijn graph, we assemble contigs using a very simple
greedy approach. A read sequence is used as a seed and extended on both ends
one nucleotide at a time by repeatedly finding the most abundant $k$-mer that
overlaps the end of the contig by $k-1$ bases. More sophisticated heuristics
have been developed, but we choose to focus on efficiency, sacrificing a
degree of accuracy.

Counting $k$-mers efficiently with the help of Bloom filters was previously
explored by \citet{Melsted2011}, who use it in addition, rather than
in place, of a hash table. The Bloom filter is used as a ``staging area'' to
store $k$-mers occurring only once, reducing the memory required by the hash
table. Concurrently with our work, \citet{Pell2011} have also developed a
probabilistic de Bruijn graph assembler, but using a non-counting Bloom
filter. While they demonstrate a significant reduction in memory use, unlike
other de Bruijn graph assemblers, only the presence or absence of a k-mer is
stored, not its abundance, which is essential information when the goal is
producing accurate contigs.

\subsection{Metadata}

In designing the file format used by Quip, we included several useful features
to protect data integrity. First, output is divided into blocks of several
magabytes each. In each block a separate 64 bit checksum is computed for read
identifiers, nucleotide sequences, and quality scores. When the archive is
decompressed, these checksums are recomputed on the decompressed data and
compared to the stored checksums, verifying the correctness of the output. The
integrity of an archived dataset can also be checked with the ``quip --test''
command.

Apart from data corruption, reference-based compression creates the
possibility of data loss if the reference used for compression is lost, or an
incorrect reference is used. To protect against this, Quip files store a 64
bit hash of the reference sequence, ensuring that the same sequence is used
for decompression. To assist in locating the correct reference, the file name,
and the lengths and names of the sequences used in compression are also stored
and accessible without decompression.

Additionally, block headers store the number of reads and bases compressed in
the block, allowing summary statistics of a dataset to be listed without
decompression using the ``quip --list'' command.

\section{Results}

\subsection{Compression of Sequencing Data}


We compared Quip to three commonplace general-purpose compression algorithms:
gzip, bzip2, and xz, as well as more recently developed domain-specific
compression algorithms: DSRC \citep{Deorowicz2011} and Cramtools
\citep{Hsi-YangFritz2011}. Other methods have been proposed \citep{Tembe2010,
Kozanitis2011, Bhola2011}, but without publicly available software we were
unable independently evaluate them. We have also restricted our focus to
lossless compression, and have not evaluated a number of promising lossy
methods \citep{Hsi-YangFritz2011,Wan2011}, nor methods only capable
of compressing nucleotide sequences \citep{Cox2012}. We invoked Quip in three
ways: using only statistical compression (``quip''), using reference-based
compression (``quip -r''), and finally with assembly-based compression (``quip
-a''). Table \ref{tab:methods} gives an overview of the methods evaluated.

We acquired six datasets (Table \ref{tab:datasets}) from the Sequence Read
Archive \citep{Leinonen2011}, representing a broad sampling of recent
applications of next-generation sequencing. Genome and exome sequencing data was
taken from 1000 Genomes Project
\citep{The1000GenomesConsortium2010},
total and mRNA data from the Illumina BodyMap
2.0 dataset \citep{Asmann2012},
Runx2 ChIP-Seq performed in a prostate cancer cell line 
\citep{Little2011}, and metagenomic DNA sequencing of biofilm found in the extremely acidic
drainage from the Richmond Mine in California
\citep{Denef2010}.

The Sequence Read Archive provides data in its own SRA compression format
which we also evaluate here. Programs for working with SRA files are provided
in the SRA Toolkit, but a convenient means of converting from FASTQ to SRA is
not, so we have not measured compression time and memory in this case.


In the case of reference-based compression implemented in Cramtools and Quip, we
aligned reads from all but the metagenomic data to the GRCh37/hg19 assembly of
the human genome using GSNAP \citep{Wu2010} generating a BAM file.
Splice-junction prediction was enabled for the two RNA-Seq datasets. In the case of
multiple alignments, we removed all but the the primary (i.e., highest
scoring). Quip is able to store secondary alignments, but the version of
Cramtools we evaluated was not. When the purpose of alignment in simply
compactly archiving the the reads, secondary alignments have no purpose and
merely inflate the file size, but in downstream analysis they are often
extremely informative and retaining them may be desirable in some cases.


It is important to note pure lossless compression is not the intended goal of
Cramtools. Though we invoked Cramtools with options to include all quality
scores (\texttt{--capture-all-quality-scores}), optional tags (\texttt
{--capture-all-tags}), and retain unaligned reads (
\texttt{--include-unmapped-reads}), it is unable to store the original read IDs. This puts it at
advantage when comparing compressed file size, as all other programs compared
were entirely lossless, but as the only other available reference-based
compression method, it is a useful comparison.

\begin{figure*}
\begin{center}
\includegraphics[width=\textwidth]{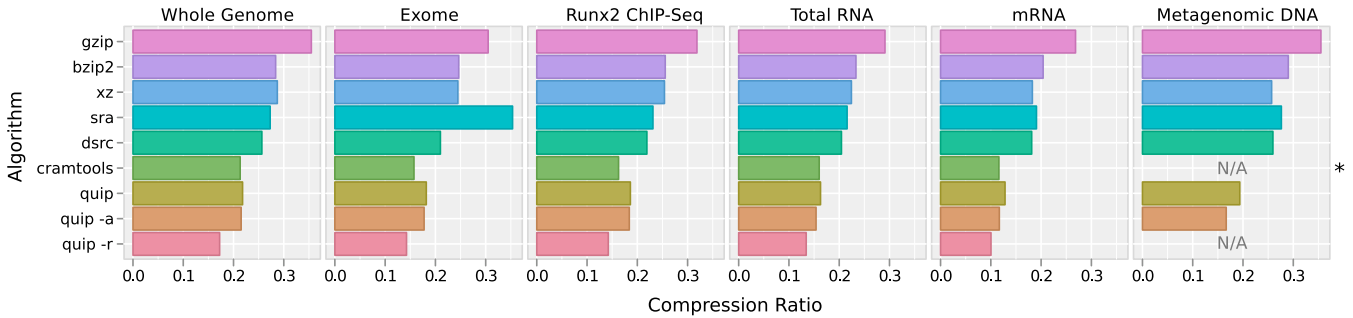}
\end{center}
\caption{
One lane of sequencing data from each of six publicly available datasets
(Table \ref{tab:datasets}) was compressed using a variety of methods (Table
\ref{tab:methods}). The size of the compressed data is plotted in proportion
to the size of uncompressed data. Note that all methods evaluated are entirely
lossless with the exception of Cramtools, marked in this plot with an
asterisk, which does not preserve read IDs, giving it some advantage in these
comparisons. Reference-based compression methods, in which an external
sequence database is used, were not applied to the metagenomic dataset for lack of an
obvious reference. These
plots are marked ``N/A''.
}
\label{fig:sizes}
\end{figure*}
\begin{figure*}
\centerline{\includegraphics[width=\textwidth]{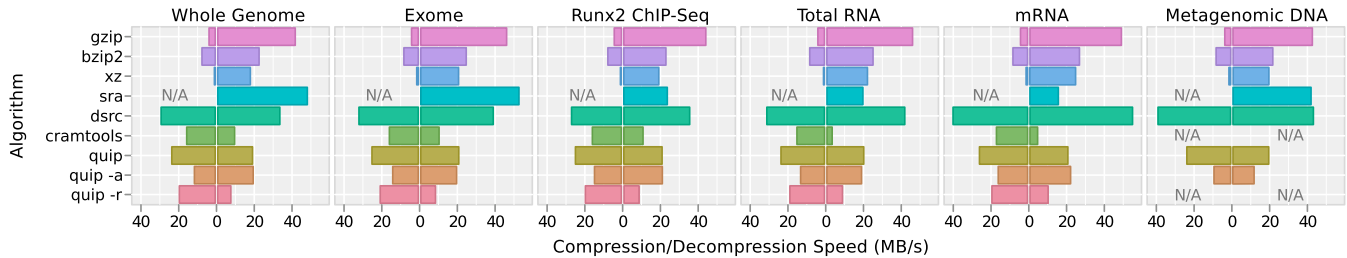}}
\caption{
The wall clock run-time of each evaluation was recorded using the Unix time
command. Run-time is normalized by the size of the original FASTQ file to
obtain a measure of compression and decompression throughput, plotted here in megabytes
per second. Compression speed is plotted to the left of the zero axis, and
decompression speed to the right. There is a not a convenient means of
converting from FASTQ to SRA, so compression time is not measured for SRA.
Additionally, reference-based methods are not applied to the metagenomic data.
In the case of reference-based compression implemented in ``cramtools'' and
``quip -r'', input and output is in the BAM format, so the speeds plotted here
include time needed to decompress and compress the BAM input and output,
respectively.
}
\label{fig:comp_decomp_time}
\end{figure*}

With each method we compressed then decompressed each dataset on a server with
a 2.8Ghz Intel Xeon processor and 64 GB of memory. In addition to recording
the compressed file size (Figure \ref{fig:sizes}), we measured wall clock
run-time (Figure \ref{fig:comp_decomp_time}) and maximum memory usage (Table
\ref{tab:mem_usage}) using the Unix \texttt{time} command. Run time was 
normalized to the size of the original FASTQ file, measuring throughput in
megabytes per second.

For the reference-based compressors, time spent aligning reads to the genome
was not included, though it required up to several hours. Furthermore, while
Quip is able to compress aligned reads in an uncompressed SAM file and then
decompress back to SAM (or FASTQ), the input and output of Cramtools is
necessarily a BAM file, and so we also use BAM files as input and output in
our benchmarks of ``quip -r''. This conflates measurements of compression and
decompression time: since BAM files are compressed with zlib, compression
times for the reference-based compressors include time needed to decompress
the BAM file, and decompression times include re-compressing the output to
BAM. Consequently, decompression speeds in particular for both Cramtools and
reference-based Quip (``quip -r'') appear significantly lower than they
otherwise might.

In the single-genome samples, we find that reference-based compression using
Quip consistently results in the smallest file size (i.e., highest
compression), yet even without a reference sequence, compression is quite high,
even matching the reference-based approach used by Cramtools in two of the
datasets. Assembly-based compression provides almost no advantage in the
single-genome ChIP-Seq, Whole Genome, and Exome datasets. Because we assembly
only several million reads (2.5 million, by default), coverage is only enough
to assemble representatives of certain families of repeats in these datasets.

In the RNA-Seq datasets, more is assembled (approximately 35\% and 70\% of the
reads are aligned to assembled contigs in the total RNA-Seq and mRNA-Seq
datasets, respectively), but this only provides a small increase in the
compression overall. Mostly probably, the aligned reads in these datasets are
relatively low-entropy, composed largely of protein-coding sequence, and would
be encoded only moderately less compactly with the Markov chain method.

Nevertheless, assembly-based compression does provide a significant advantage
in the metagenomic dataset. Nearly 95\% of reads align to assembled contigs,
resulting in a large reduction in compressed file size. Though this may be the
result of low species diversity and may not extend to all metagenomic
datasets, it shows that assembly-based compression is beneficial in certain
cases. Though it comes at some cost, we are able to perform assembly with
exceedingly  high efficiently: memory usage is increased only by 400 MB and
compression time is faster than all but DSRC and assembly-free Quip even while
assembling 2.5 million reads and aligning millions more reads to the assembled
contigs.

\begin{figure*}
\centerline{\includegraphics[width=\textwidth]{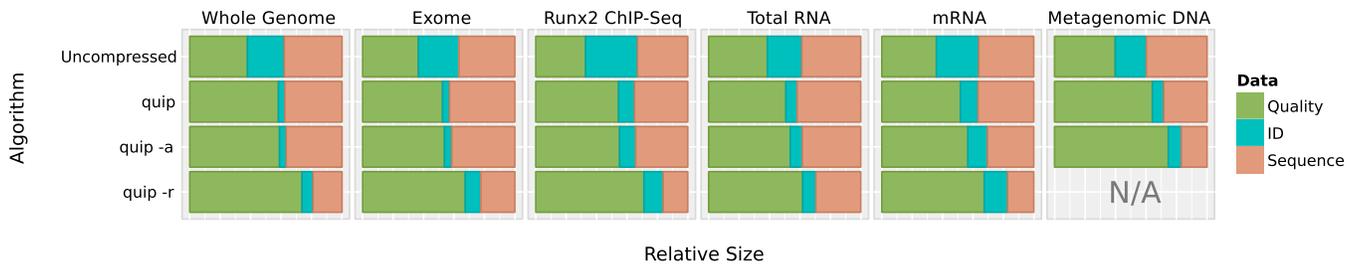}}
\caption{
After compressing with Quip using the base algorithm (labeled ``quip''),
assembly-based compression (``quip -a''), and reference-based compression
(``quip -r''), we measure the size of read identifiers, nucleotide sequences,
and quality scores in proportion to the total compressed file size. Reference-based
compression (``quip -r'') was not applied to the metagenomic data.
}
\label{fig:relative_sizes}
\end{figure*}

With all invocations of Quip, the size of the resulting file is dominated,
sometimes overwhelmingly, by quality scores (Figure \ref{fig:relative_sizes}).
There is not a direct way to measure this using the other methods evaluated,
but this result suggests that increased compression of nucleotide sequences
will result in diminishing returns.

\begin{table*}
\begin{tabular}{@{}lll@{}}
Program     & Input & Methods \\
gzip (1.4)           & Any &
Lempel-Ziv \\
bzip2 (1.0.6)        & Any &
Burrows-Wheeler and Huffman coding \\
xz  (5.0.3)          & Any & 
Lempel-Ziv with arithmetic coding \\
sra (2.1.10)         & FASTQ &
No published description \\
dsrc (0.3)           & FASTQ &
Lempel-Ziv and Huffman coding. \\
cramtools (0.85-b51) & BAM &
Reference-based compression \\
quip (1.1.0)         & FASTQ, SAM, BAM &
Statistical modeling with arithmetic coding \\
quip -a (1.1.0)      & FASTQ, SAM, BAM &
Quip with assembly-based compression \\
quip -r (1.1.0) & SAM, BAM &
Quip with reference-reference based compression \\
\end{tabular}
\caption{Methods evaluated in the results section.
All methods compared are lossless, with the exception of Cramtools which
does not preserve the original read identifiers.}
\label{tab:methods}
\end{table*}

\begin{table*}
\begin{tabular}{@{}lllrr@{}}
Accession Number & Description     & Source & Read Length & Size (GB) \\
SRR400039        & Whole Genome    &
The 1000 Genomes Project \citep{The1000GenomesConsortium2010} & 101 & 67.6 \\
SRR125858        & Exome           &
The 1000 Genomes Project \citep{The1000GenomesConsortium2010} & 76 & 55.2 \\
SRR372816        & Runx2 ChIP-Seq  & 
Little, et al \citep{Little2011}
&          50 & 14.6 \\
ERR030867        & Total RNA       &
The BodyMap 2.0 Project, Asmann, et al \citep{Asmann2012}
&         100 & 19.5 \\
ERR030894        & mRNA            &  
The BodyMap 2.0 Project, Asmann, et al \citep{Asmann2012}
&          75 & 16.4 \\
SRR359032        & Metagenomic DNA & 
Denef, et al \citep{Denef2010}
&         100 &  8.9 \\
\end{tabular}
\caption{
We evaluated compression on a single lane of sequencing data taken from a
broad collection of seven studies. Except for the metagenomic data, each was
generated from human samples. Uncompressed file sizes are shown in gigabytes.
}
\label{tab:datasets}
\end{table*}

\begin{table}[h]
\begin{tabular}{@{}lrr@{}}
                & \multicolumn{2}{c}{\textit{Average Memory Usage (MB)}} \\ \cline{2-3}
\textit{Method} & \textit{Compression} & \textit{Decompression} \\
gzip            &     0.8              &    0.7 \\
bzip2           &     7.0              &    4.0 \\
xz              &    96.2              &    9.0 \\
sra             &      NA              &   46.9 \\
dsrc            &    28.3              &   15.1 \\
cramtools       &  6803.8              & 6749.3 \\
quip            &   412.1              &  412.9 \\
quip -a         &   823.0              &  794.1 \\
quip -r         &  1710.0              & 1717.2 \\
\end{tabular}
\caption{
Maximum memory usage was recorded for each program and dataset. We list
the average across datasets of these measurements. For most of the programs,
memory usage varied by less that 10\% across datasets and well summarized by the mean,
with the exception of SRA. Decompression used less that 60 MB except in the
human exome sequencing dataset, in which over 1.4 GB was used. We report the mean
with this outlier removed.}
\label{tab:mem_usage}
\end{table}

\subsection{Characteristics of the d-left Counting Bloom Filter}

Though our primary goal is efficient compression of sequencing data, the
assembly algorithm we developed to achieve this is of independent interest.
Only very recently has the idea of using probabilistic data structures in
assembly been breached, and to our knowledge, we are the first to build a
functioning assembler using any version of the counting Bloom filter. Data
structures based on the Bloom filter make a trade-off between space and
the probability of inserted elements colliding. Tables can be made arbitrarily
small at the cost of increasing the number of collisions. To elucidate this
trade-off we performed a simple simulation comparing the dlCBF to a hash table.

Comparisons of data structures are notoriously sensitive to the specifics of
the implementation. To perform a meaningful benchmark, we compared our dlCBF
implementation to the sparsehash library
(\url{http://code.google.com/p/sparsehash/}), an open-source hash table
implementation with the expressed goal of maximizing space efficiency. Among
many other uses, it is the core data structure in the ABySS
\citep{Simpson2011} and PASHA \citep{Liu2011} assemblers.

We randomly generated 10 million unique, uniformly distributed 25-mers and
inserted them into a hash table, allocated in advance to be of minimal size
while avoiding resizing and rehashing. We repeated this with dlCBF tables of
increasing sizes. Upon insertion, a collision occurs when the hash functions
computed on the inserted $k$-mer collide with a previously inserted $k$-mer.
An insertion may also fail when a fixed size table is filled to capacity and
no empty cells are available. For simplicity, we count these occurrences also
as collisions. Wall clock run-time and maximum memory usage were both recorded
using the Unix time command.

\begin{figure}[h]
\centerline{\includegraphics[width=\columnwidth]{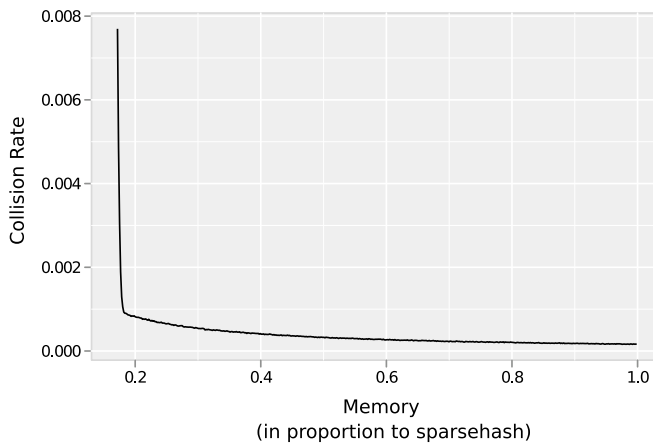}}
\caption{
The trade-off between memory usage and false positive rate in the dlCBF is
evaluated by inserting 10 million unique 25-mers into tables of increasing
size. Memory usage is reported as the proportion of the memory used by a
memory efficient hash table to do the same.
}
\label{fig:dlcbf_bench}
\end{figure}

We find that with only 20\% of the space of the hash table the dlCBF accrues a
collision rate of less than 0.001 (Figure \ref{fig:dlcbf_bench}). While
the hash table performed the 10 million insertions in 7.34 seconds, it
required only 4.48 seconds on average for the dlCBF to do the same, with both
simulations run on a 2.8Ghz Intel Xeon processor. Table size did not greatly
affect the runtime of the dlCBF. The assembler implemented in Quip uses a
fixed table size, conservatively set to allow for 100 million unique $k$-mers
with a collision rate of approximately 0.08\% in simulations.

Though the authors of sparsehash claim only a 4 bit overhead for each entry,
and have gone to considerably effort to achieve such efficiency, it still must
store the $k$-mer itself, encoded in 64 bits. The dlCBF avoids this, storing
instead a 14-bit ``fingerprint'', or hash of a $k$-mer, resulting in the large
savings we observe. Of course, a 25-mer cannot be uniquely identified with 14
bits. False positives are thus introduced, yet they are kept at a very low
rate by the d-left hashing scheme. Multiple hash functions are used under this
scheme, so that multiple hash function must collide to result in a
collision, an infrequent event if reasonably high-quality hash functions are
chosen.

A previous analysis of the dlCBF by \citet{Zhang2009} compared it to two other
variations of the counting Bloom filter and concluded that ``the dlCBF
outperforms the others remarkably, in terms of both space efficiency and
accuracy.'' Overall, this data structure appears particularly adept for high
efficiency de novo assembly.

\section{Discussion}

Compared to the only other published and freely-available reference-based
compressor, Cramtools, we see a significant reduction in compressed file size,
despite read identifiers being discarded by Cramtools and retained by Quip.
This is combined with dramatically reduced memory usage, comparable run-time
(slightly faster compression paired with varying relative decompression
speeds), and increased flexibility (e.g. Quip is able to store multiple
alignments of the same read and does not require that the BAM file be sorted
or indexed). Conversely, Cramtools implements some potentially very useful
lossy methods not provided by Quip. For example, a number of options are
available for selectively discarding quality scores that are deemed
unnecessary, which can significantly reduce compressed file sizes.

We find that assembly-based compression offers only a small advantage in many
datasets. Because we assemble a relatively small subset of the data, coverage
is too low when sequencing a large portion of a eukaryotic genome. In
principle, the assembler used could be scaled to handle exponentially more
reads, but at the cost of being increasingly computationally intensive. In
such cases, the reference-based approach may be more appropriate. Yet, in the
metagenomic dataset we evaluate, where there is no obvious reference sequence,
nucleotide sequences are reduced in size by over 40\% (Figure
\ref{fig:relative_sizes}). While limited in scope, for certain projects
assembly-based compression can prove to be invaluable, and the algorithm devised
here makes it a tractable option.

The Lempel-Ziv algorithm, particularly as implemented in gzip/libz has become
a reasonable default choice for compression. The zlib library has matured and
stabilized over the course of two decades and is widely available. The BAM and
Goby formats both use zlib for compression, and compressing FASTQ files with
gzip is still common practice. Despite its ubiquity, our benchmarks show that
it is remarkably poorly suited to NGS data. Both compression ratio and
compression time were inferior to the other programs evaluated. For most
purposes, the gains in decompression time do not make up for its shortcomings.
With the more sophisticated variation implemented in xz, compression is improved
but at the cost of tremendously slow compression.

Our use of high-order Markov chain models and de novo assembly results in a
program that uses significantly more memory than the others tested, with the
exception of Cramtools. Though limiting the memory used by a general purpose
compression program enables it to be used on a wider variety of systems, this
is less important in this domain-specific application. Common analysis of
next-generation sequencing data, whether it be alignment, assembly, isoform
quantification, peak calling, or SNP calling all require significant
computational resources. Targeting low-memory systems would not be of
particular benefit: next-generation sequencing precludes previous-generation
hardware.

Though memory consumption is not a top priority, runtime is important. Newer
instruments like the HiSeq 2000 produce far more data per lane than previously
possible. And, as the cost of sequencing continues to decrease, experiments
will involve more conditions, replicates, and timepoints.  Quip is able to
compress NGS data at three times the speed of gzip, while performing de novo
assembly of millions of reads, and up to five times as fast without the
assembly step. Only DSRC is faster, but with consistently lower compression.
In addition, our reference-based compression algorithm matches the speed
of cramtools but with substantially better lossless compression.

As illustrated in Figure \ref{fig:relative_sizes}, the size of compressed NGS
data is dominated by quality scores. More sophisticated compression of
nucleotide sequences may be possible but will result in only minor reductions
to the overall file size. The largest benefit would be seen by reducing
quality scores. While it is easy to reduce the size of quality data by coarse
binning or other lossy methods, it is very hard to determine the effect such
transformations will have on downstream analysis. Future work should
concentrate on studying lossy quality score compression, strictly guided by
minimizing loss of accuracy in alignment, SNP calling, and other applications.
Quality scores are encoded in Quip using an algorithm that is suitably
general so that lossy transformations (e.g., binning) can be automatically
exploited, and can be treated as an optional preprocessing step.

Other aspects of the algorithm presented here can be improved, and will be
with future versions of the software. A large body of work exists exploring
heuristics to improve the the quality of de novo assembly, however the
algorithm in Quip uses a very simple greedy approach. Assembly could likely be
improved without greatly reducing efficiency by exploring more sophisticated
methods. We currently perform no special handling of paired-end reads, so that
mates are compressed independently of each other. In principle some gains to
compression could be made by exploiting pair information. We also intend to
implement parallel compression and decompression. This is non-trivial,
but quite possible and worthwhile given the abundance of data and ubiquity of
multi-core processors.

Combining reference-based and assembly-based techniques, with carefully tuned
statistical compression, the algorithm presented in Quip probes the limit to
which next-generation sequencing data can be compressed losslessly, yet
remains efficient enough be a practical tool when coping with the deluge of
data that biology research is now presented with.

\section{Funding}

Federal funds from the National Institute of Allergy and Infectious Diseases,
National Institutes of Health, Department of Health and Human Services, under
Contract No. HHSN272200800060C and by the Public Health Service grant
(P51RR000166) from the National Institutes of Health, in whole or in part.

\section{Acknowledgments}

We are grateful to James Bonfield, Matt Mahoney, Seth Hillbrand and other
participants in the Pistoia Alliance Sequence Squeeze competition, whose work
inspired many improvements to our own. While preparing the manuscript, Richard
Green provided valuable feedback and suggestions.

\subsubsection{Conflict of interest statement.} None declared.
\newpage

\bibliographystyle{nar}
\bibliography{quip}

\end{document}